\documentclass[]{emulateapj} % <-- emulador de l'ApJ
\usepackage{apjfonts}                   %     publicat

\usepackage{epsfig,graphicx,latexsym,amsmath,amssymb}
\usepackage{hyperref}
\usepackage{mathrsfs}
\usepackage{lastpage}

\usepackage{booktabs}

\usepackage{color}
\usepackage{natbib}

\newcommand{\fb}{{\sc Fewbody }}
\newcommand{\PN}{{\rm{PN}}}

\bibpunct[,]{(}{)}{;}{a}{}{,}

\begin{document}

\author{
Pau Amaro-Seoane\altaffilmark{1}\thanks{e-mail: Pau.Amaro-Seoane@aei.mpg.de} \&
Xian Chen\altaffilmark{2}\thanks{e-mail: xchen@astro.puc.cl}
}

\altaffiltext{1}{Max Planck Institut f\"ur Gravitationsphysik
(Albert-Einstein-Institut), D-14476 Potsdam, Germany}

\altaffiltext{2}{xchen@astro.puc.cl, Instituto de Astrof\'{i}sica, Facultad de
F\'{i}sica, Pontificia Universidad Cat\'{o}lica de Chile, 782-0436 Santiago,
Chile}

\date{\today}

\label{firstpage}

\title{Relativistic mergers of black hole binaries\\
       have large, similar masses, low spins and are circular}

\begin{abstract}

Gravitational waves are a prediction of general relativity, and with
ground-based detectors now running in their advanced configuration, we will
soon be able to measure them directly for the first time.  Binaries of
stellar-mass black holes are among the most interesting sources for these
detectors.  Unfortunately, the many different parameters associated with the
problem make it difficult to promptly produce a large set of waveforms for the
search in the data stream.  To reduce the number of templates to develop, one
must restrict some of the physical parameters to a certain range of values
predicted by either (electromagnetic) observations or theoretical modeling.  In
this work we show that ``hyperstellar'' black holes (HSBs) with masses $30
\lesssim M_{\rm BH}/M_{\odot} \lesssim 100$, i.e black holes significantly
larger than the nominal $10\,M_{\odot}$, will have an associated low value for
the spin, i.e. $a<0.5$. We prove that this is true regardless of the formation
channel, and that when two HSBs build a binary, each of the spin magnitudes is
also low, and the binary members have similar masses. We also address the
distribution of the eccentricities of HSB binaries in dense stellar systems
using a large suite of three-body scattering experiments that include
binary-single interactions and long-lived hierarchical systems with a highly
accurate integrator, including relativistic corrections up to ${\cal
O}(1/c^5)$. We find that most sources in the detector band will have nearly
zero eccentricities.  This correlation between large, similar masses, low spin
and low eccentricity will help to accelerate the searches for
gravitational-wave signals.

\end{abstract}


\maketitle

\section{Introduction}

The first-generation ground-based detector Laser Interferometer
Gravitational-wave Observatory (LIGO) has successfully undergone major
technical upgrades in the past few years that have led to a significant increase of the
volume of the observable universe
\footnote{\scriptsize{http://www.ligo.caltech.edu/advLIGO/ }}.  The detector
has already been operative for a few months and \textit{observation run 1} (O1)
is being analyzed as these lines are written.  In this configuration, LIGO can
observe binaries of stellar-mass black holes (BHs)---one of the most
interesting sources to be detected---of masses $(25+25)\,M_{\odot}$ out to a
distance of $\sim 3.4$ Gpc (a redshift of $z\sim0.2-0.3$), and with the final
advanced (aLIGO) configuration about a factor 3 farther away (see
\citealt{BrownEtAl2013} and \citealt{2013arXiv1304.0670L}).  On the other hand,
the Virgo Interferometer\footnote{\url{http://www.virgo-gw.eu/}} is currently
undergoing an upgrade program to improve its strain sensitivity, and will be
online in the advanced configuration in the next future.
The first direct detection of GWs is imminent{, provided the number of sources in the observable volume of aLIGO is big enough}.

The fundamental low-frequency limitations of the second-generation detectors
are given by thermal, gravity gradient, and seismic noise. To circumvent these
problems, yet a third generation of gravitational-wave (GW) interferometers to
be operated underground is currently being proposed. The Einstein
Telescope\footnote{\url{http://www.et-gw.eu/}} will be a 10 km
laser-interferometer with a sensitivity 100 times better than that of the
current detectors, which expands the observable volume of the universe by a
factor of a million \citep{SathyaprakashSchutzBroeck09}. Moreover it will cover
the frequency range between 1~Hz and $10^4$~Hz.

For the search, the availability of accurate waveform models for the full
merger is pivotal.  Numerical relativity succeeded ten years ago in simulating
the late inspiral, merger and ringdown
\citep{Pretorius:2005gq,Campanelli:2005dd,Baker05a}. Together with
post-Newtonian modeling of the inspiral phase, data analysts can do faithful
searches for binaries of BHs of comparable-mass and with mass ratios up to
about 10 \citep[see
e.g.][]{Buonanno:1998gg,BuonannoEtAl07,Ajith:2009bn,SantamariaEtAl10}.
Nonetheless, the high cost of the development of the many ($\sim\,10^5-10^6$)
waveforms necessary for a
matched-filtering search represents a problem, if not a true limitation.

Hence, to speed up the production of waveforms, certain values are assumed for
some of the physical parameters.  These values are not chosen randomly, but
rely on electromagnetic observations or theoretical modeling, which represents
an emergent symbiotic relationship between astrophysics and GW searches---we reduce
the spectrum of the parameter space based on our best understanding of the
astrophysics of the system and, once the first detections arrive, they will
help us to understand our astrophysical models better.

Following this line of thought, in this paper we show that there is a
correlation between the mass of BHs with large masses, which we call
``hyperstellar'' (henceforth HSB)\footnote { Depending on the mass, a BH is (a)
supermassive ($\gtrsim 10^6\,M_{\odot}$), (b) stellar-mass ($\sim
10\,M_{\odot}$) or (c) intermediate-massive (IMBH,
$\left[100,\,10^5\right]\,M_{\odot}$). The range of masses we are interested in
is $20 \lesssim M_{\rm BH}/M_{\odot} \lesssim 100$. Since these BHs are
too heavy compared to the nominal mass of an stellar-mass BH but well below the
mass of an IMBH, we choose the name ``hyperstellar'' to avoid confusion.  } ,
and their spin. HSBs are expected to form in low-metallicity environments
following the collapse of very-massive stars, as in the works of \cite{woosley02,heger03}, and have been suggested to be the
central engines of those ultra-luminous X-ray sources found in star-forming galaxies
(\citealt{mapelli09,mapelli10,mapelli14}, who call this sort of BHs ``massive stellar black holes'').
\textit{We show that BHs with masses}
$\,\gtrsim\,30\,M_{\odot}$ \textit{will be detected with typically low spin
values} ($\lesssim\,0.50$) \textit{and on basically circular binary orbits}.

\section{Binaries of hyperstellar black holes}\label{sec:binary}

Binaries of compact objects in the relativistic regime detectable by ground-based
interferometers like aLIGO can, in principle, form in two different ways (see
e.g. the Living Review paper by \citealt{BenacquistaDowning13} and references
therein): (i) in galactic plane or bulge as the remnants of massive binary
stars (``field binaries''), or (ii) in dense stellar-systems, such as
globular and young clusters, or those nuclear star clusters at the centers
of galaxies, via dynamical interactions (henceforth ``dynamical binaries'').

\subsection{Field binaries}

To have two BHs in a field binary, we need a binary formed of two stars that
are both massive enough \citep[e.g.][]{demink15}. Since we need them bound, the
natal kick that the BHs receive during their formation cannot exceed the
break-up velocity of the stellar binary. This favors the formation of binaries
of HSBs because (i) high-mass BHs receive relatively small natal kicks
\citep{belczynski15} and (ii) a higher mass requires a larger break-up
velocity. For these reasons, we expect the ground-based GW experiments to
detect mostly HSBs \citep[see][~for a more quantitative
evaluation]{dominik15}.

\subsection{Dynamical binaries}

Dynamical binaries will naturally tend to form with large masses. This is so
because the timescale for an object to sink towards the center of a stellar
system via dynamical friction is the relaxation time divided by its mass --
more massive stars will sink first into a dense environment.  Moreover,
binaries of larger masses are more difficult to separate in three-body
interactions because of the larger binding energy.

In particular, \cite{oleary06} found with Monte-Carlo simulations BH masses
well exceeding the nominal $10\,M_{\odot}$ in $(20-80)\%$ of the binaries in
globular clusters, and \cite{MillerLauburg09} estimated with semi-analytical
arguments and scattering experiments that there is a strong tendency for the
merging BHs in nuclear star clusters to be biased toward high masses.

This bias of forming massive binaries due to the existence of HSBs has been
confirmed using direct-summation $N-$body simulations of young clusters
\citep{mapelli13}.
As a result of the above bias, ground-based observatories are more likely
to detect HSB binaries in dense clusters than in the stellar-mass range (i.e. with masses of about $10M_{\odot}$ \citealt{OLearyEtAl06,MillerLauburg09,ziosi14,rodriguez15}).

\section{Estimation of the spin for field HSB binaries}
\label{sec:field}

In the field, the only way to produce a HSB binary is to form, in the first
place, a stellar binary of two Wolf-Rayet (WR) stars \citep[with stellar mass
$M_*>60~M_\odot$, following the definition of][]{zinnecker07}.  An environment
with low metallicity is more favorable. For example, in a solar-metallicity
environment, a $80~M_\odot$ WR star produces a BH of only $10~M_\odot$
\citep{woosley02}, but in an environment with $0.1$ solar metallicity the same
star can produce a $30~M_\odot$ HSB \citep[e.g.][]{dominik15}.  It is
interesting to note that for extremely metal-poor environments (which are
uncommon in the redshift range of our interest, $z\la0.2-0.3$,
\citealt{panter08}), HSBs in the mass range $25-55\,M_\odot$ are prevented from
forming because of a pair-instability process during the supernova phase
\citep{heger03,fryer12}.

Because of angular-momentum loss by stellar wind, WR stars are slow rotators.
The rotational velocity at the stellar surface drops from an initial value of
$200-300~{\rm km~s^{-1}}$ to below $50~{\rm km~s^{-1}}$ during the stellar
evolution \citep{meynet03,meynet05}.  As a result, the compact remnant which is
sitting at the core of a WR star retains very little of the initial angular
momentum.  For example, when $M_*=80~M_\odot$---which is the initial stellar
mass relevant to HSB formation---the current stellar evolution models predict
that the specific angular momentum measured at the edge of the stellar remnant
will drop from $\sim10^{18}~{\rm cm^2~s^{-1}}$ in the initial configuration to
as small as $j_{\rm rem}\simeq(1-6)\times10^{16}~{\rm cm^2~s^{-1}}$, and even
smaller for more massive stars \citep{hirschi05,yusof13}.

Since the remnant of a WR star has a low specific angular momentum, the HSB
must be a slow rotator.  Even under the assumption that no angular momentum is
lost during the collapse of the remnant and the growing of the initial seed BH
that eventually becomes a HSB, the spin parameter of the final HSB is:

\begin{eqnarray}
a=\frac{j_{\rm rem}M_\bullet}{GM^2_\bullet/c}
\simeq(0.075-0.45)\left(\frac{M_\bullet}{30~M_\odot}\right)^{-1},
\label{eq.spin_param}
\end{eqnarray}
where $c$ is the speed of light. Note that because of the reason we just
argued, this is an upper limit, and in a more realistic estimation, the value
will be even lower.  Also, it is interesting to note that
Equation~(\ref{eq.spin_param}) gives the same spin range as what we have
observed currently for the stellar-mass BHs ($M_\bullet<20~M_\odot$) in the
Local Group \citep{mcclintock14}.

To spin up the slowly rotating HSB that we have predicted, the only two
possibilities are (i) merger with another HSB because of a dynamical
interaction, such as a three-body encounter, or (ii) accretion of enough
material onto the HSB, about an amount of $M_\bullet$ to achieve a result of
$a>0.5$. The first possibility can indeed enhance the spin value of the merger
product to $a>0.5$. However, this process cannot take place in the field,
because the stellar relaxation timescale (i.e. encounter timescale) exceeds by
many orders of magnitude the Hubble time.  Moreover, in the next section we
will see that the outcome will not be observed because of a dynamical selection
effect.

Regarding (ii), there are two sources that can possibly provide material to
feed the HSB, either the interstellar medium (ISM) surrounding our field
binary, or the companion star, i.e. the other WR star in the binary which may
not have collapsed into a BH yet.  The Bondi-Hoyle accretion rate from the ISM
is $10^{-15}~M_\odot~{\rm yr^{-1}}(\sigma/10^2~{\rm km~s^{-1}})(n_{\rm
ISM}/{1\cdot \rm cm^{3}})(M_\bullet/30~M_\odot)^2$ where $n_{\rm ISM}$ is the
proton density in ISM and $\sigma$ the velocity dispersion of stars
\citep{MillerLauburg09}. This feeding rate is too low to spin up the HSB,
because it would require a time longer than the Hubble time. On the other hand,
the life span of the WR companion star is of $5$ Myr, which is too short
compared to the minimum timescale for a BH to double its mass by accreting at
the maximum (Eddington) rate . This is the Salpeter timescale, which is about
$30$ Myr assuming that $10\%$ of the rest mass is converted into radiation
{ \citep[see Sec. 6.3 in][]{mcclintock14}}.

We hence deduce that \textit{field HSBs must be slow rotators, with lower
values than the upper limit shown in Eq.~\ref{eq.spin_param} because of loss of
angular momentum during the remnant collapse}.

\section{Estimation of the spin for dynamical HSB binaries}
\label{sec:cluster}

For this section, we assume that the HSB binary forms dynamically from two HSBs
that were born isolated (if this is not the case, then the previous result
applies) and then later formed a binary via three-body interactions.

There are three possible channels to form an isolated HSB: (i) The collapse of
an isolated WR star, (ii) the coalescence of two less massive BHs (e.g.
$10<(M_\bullet/M_\odot)<30$) and (iii) by accreting background stars.

Channel (i): As we have seen in Section~\ref{sec:field}, HSBs with masses
higher than $30~M_\odot$ have typically low spins, $a<0.5$.  This HSB receives
a natal kick typically of $15[M_\bullet/(30~M_\odot)]^{-1}~{\rm km~s^{-1}}$
\citep{belczynski15} and is hence kept in the host cluster, because the
velocity dispersion is of order $10~{\rm km~s^{-1}}$. Lighter BHs will not be
retained. Since we want to build binaries, it is important to retain them in
the cluster. Hence, {most} BHs with masses lighter than $30~M_\odot$ are
typically bound to leave the cluster at birth, without a chance of forming a
binary.

Channel (ii): Although we have just shown that lighter BHs typically abandon
the host cluster at birth, there is { sufficient margin} so that some remain (see e.g. the
work of \citealt{StraderEtAl2012}, although the accretion rate is so low that
it is difficult to assess whether the BHs are $10$ or $20\,M_{\odot}$), and so
we also address this channel.  We point out two particular properties of
dynamical binaries that will be important in determining the spin parameter,
$a_{\rm fin}$, of the final merged BH.  (a) On average, the two members of a
dynamical BH binary very likely will have similar masses because the three-body
interaction tends to retain equal-mass binaries. This is particularly true for
BHs, since they belong to the heaviest objects in clusters, and so they have
lower probabilities to be ionized or replaced by another interloper \citep[see
e.g.][]{MillerLauburg09}.  (b) The majority of BH-BH mergers have a
zero-eccentricity orbit, as we will see in Section~\ref{sec:ecc}.  As a result
of these two points, the binary has a significant amount of angular momentum
before it merges, and hence preferentially produces a high-spin HSB by the end
of the merger.

To prove this, we calculate the distribution of $a_{\rm fin}$ for different
mass ratios ($q:=m_2/m_1$, with $m_1$ the more massive BH), using the method
developed by one of us and presented in \cite{liu12}. The algorithm chooses a
random orientation for the spin axis of each progenitor, and then, given the
mass ratio $q$ ($q\le1$ in our model) of the two BHs, computes $a_{\rm fin}$
and the relativistic recoiling velocity, $v_k$, according to the empirical
fitting formulae derived from numerical relativity simulations
\citep{rezzolla08,baker08,vanMeter10}.

\begin{figure}
\resizebox{\hsize}{!}
          {\includegraphics[scale=1,clip]{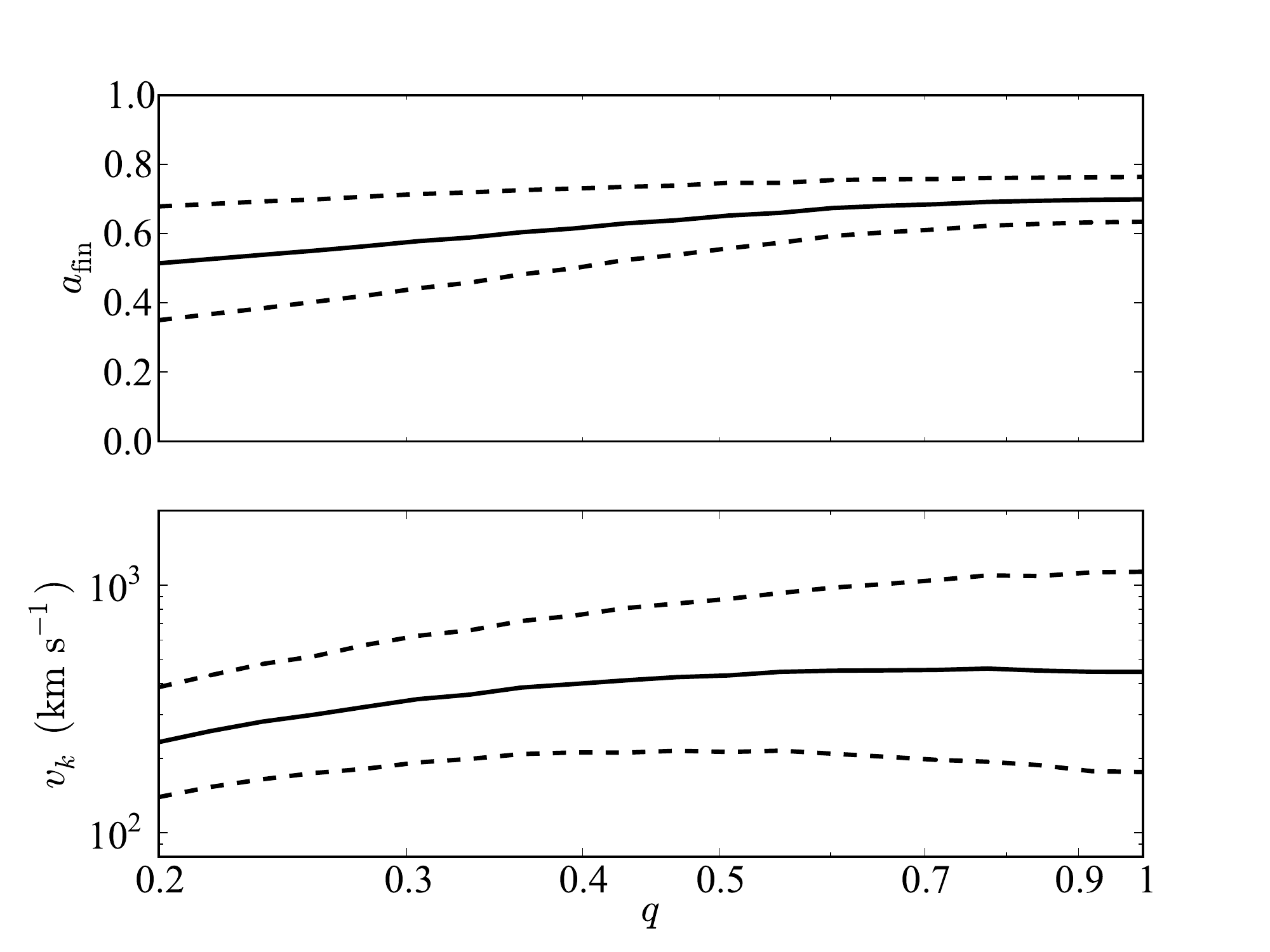}}
\caption { Distributions of the final spin $a_{\rm fin}$ (upper panel) and
recoiling velocity $v_k$ (lower panel) of a post-merger BH as a function of the
mass ratio of $q$ the two BH progenitors.  The solid lines show the mean value
and the dashed ones the $1\sigma$ range.  To derive these distributions, we run for
each $q$ value a Monte-Carlo simulation of $4000$ trials, each trial starting
with random spin magnitude (between $0$ and $0.9$) and orientation for each of
the two BH progenitors.  } \label{fig:merger} \end{figure}

The top panel of Figure~\ref{fig:merger} shows our results, where we have
assumed a random distribution for the spin magnitudes in the range $[0,\,0.9]$
for the two progenitors.  We can see that when $q>0.2$, the value of $a_{\rm
fin}$ is most likely greater than $0.4$, which correlates into a very large
relativistic recoiling velocity (see lower panel of Fig.~\ref{fig:merger}),
that, for all purposes, is much higher than the average dispersion velocity of
the host cluster, { typically of $10~{\rm km~s^{-1}}$}.  { The excess of
recoiling velocity relative to the velocity dispersion of the host stellar
system } is also true even for Nuclear Star Clusters ({ NSC}, dense stellar
clusters found at the center of galaxies of all Hubble types, see e.g.
\citealt{Boeker2010}), { whose dispersion velocities are} around $\sigma
\sim 50\,{\rm km\,s}^{-1}$, see e.g.  \citealt{MillerLauburg09}.

Hence, even if the few remaining lighter BHs led to the formation of high-spin
HSBs in clusters, these are immediately kicked out of the host environment
because of the relativistic kick, and do not stand a chance  of forming a
binary.

Channel (iii): By accreting background stars, a BH with a mass initially well
below $30~M_\odot$ cannot become a HSB. This is because the accretion rate from
other stars in a very dense environment, such as a NSC, is

\begin{align}
\Gamma&\simeq n_*\sigma r_t^2\left(1+\frac{2GM_\bullet}{r_t\sigma^2}\right) \nonumber \\
&\simeq8.6\times10^{-11}~{\rm yr^{-1}}\left(\frac{n_*}{10^6~{\rm pc^{-3}}}\right)
 \left(\frac{\sigma}{50~{\rm km/s}}\right)^{-1}
 \left(\frac{M_\bullet}{10~M_\odot}\right)^{4/3},
\label{eq.Gamma}
\end{align}
where $r_t\simeq R_\odot(M_\bullet/M_\odot)^{1/3}$ is the critical radius for
the BH to tidally disrupt a solar-type star and $n_*$ is the spatial
density of background stars. Hence, in this channel we cannot form HSBs.

Therefore HSBs retained in globular clusters or NSC
predominantly have slow spin ($a_\bullet<0.5$), and hence \textit{dynamical BH
binaries detected with ground-based detectors will be HSBs with low spin.}

\section{Eccentricity distribution}\label{sec:ecc}

Field binaries will have low eccentricities when they enter the detector band
because {their orbits circularize during the binary-star evolution due to Roche lobe overflow, common envelope phase and tidal synchronization
\citep[see e.g.][]{belczynski08,demink15} and, much later in the evolution, due to GW radiation. For dynamical binaries, however, the
eccentricities in the detector band are more uncertain, because of
dynamical interactions with other stars. For instance, a binary-binary
interaction can lead to the existence of a third body gravitationally bound
{ to and} interacting with
the binary ({ though} the probability to form a triple via single-binary interaction is very low,
at least for bodies with the same masses, see  \citealt{miller02} and references therein). For this particular situation,
the interaction can lead to a large oscillation of
the orbital eccentricity of the binary because of the Kozai-Lidov resonance \citep[see e.g.][]{blaes02}. In this case,
 as many as $50\%$ of mergers in the detector band could have an eccentricity
 larger than $e=0.1$ \citep{wen03,antonini14,antonini15}.
These results hold for bound triple systems in which all bodies have a similar or equal mass in the stellar-mass BH range (i.e. around $10\,M_{\odot}$).
However, direct-summation $N-$body simulations that do not make a priori assumptions
about mass ratios and bound systems, find that BH-BH mergers are preferentially circular ($e<0.1$) and driven by interactions with unbound small bodies (which is referred to as hardening in stellar dynamics) \citep{ziosi14}. Unfortunately, such direct-summation simulations are very expensive and not suited for statistical studies.

Hence, to statistically address the evolution of the eccentricities of the HSB
binaries during three-body interactions}, we use the numerical tool \fb
\citep{FregeauEtAl04,FregeauRappaport04}. Moreover, we have modified it by
adding relativistic corrections to the force term, which might be important in
the kind of interactions that we are interested in (HSB binaries).  For this,
we have chosen the centre-of-mass frame of the binary for the post-Newtonian ($\PN$
hereafter) expressions, which is equivalent to the center-of-mass Hamiltonian
in ADM (Arnowit, Deser and Misner) coordinates thanks to a transformation of
the particles' variables, following the expressions given by
\cite{BlanchetIyer03}. This means that the relative acceleration in the
center-of-mass frame has the form

\begin{equation}
\frac{d v^i}{dt}=-\frac{m}{r^2}\Big[(1+{\cal A})\,n^i + {\cal B}\,v^i
\Big]+ {\cal O}\left( \frac{1}{c^7} \right)\;,
\label{eq.Blanchet}
\end{equation}

\noindent
where the relative separation of the binary is $x^i=y_1^i-y_2^i$, $r=|{\bf x}|$
and $n^i={x^i}/{r}$; the ${\cal A}$ and ${\cal B}$ are given by the expressions
(3.10a) and (3.10b) in \cite{BlanchetIyer03}. We truncate the series and
neglect all terms of order higher than 2.5\,$\PN$, since e.g. the 3\,$\PN$
correction requires a very expensive computation and provides us only with a
correction which is negligible for the purpose of this study. On the other
hand, the 3.5\,$\PN$ term, even if it is significantly less challenging, would
provide us only with a rough ($\sim 10\%$) estimate of the gravitational recoil
velocity. In other words, we focus in this work only on the relativistic
corrections corresponding to periastron shift (1\,$\PN$ and 2\,$\PN$) and the
energy loss in the form of GW radiation (2.5\,$\PN$).  The first implementation
of relativistic terms in a dynamical code was done by one of us and they have
been well tested in a number of different works that feature different orders
in the expansion \citep[see
e.g.][]{KupiEtAl06,Amaro-SeoaneSopuertaBrem2012,BremAmaro-SeoaneSpurzem2013,BremAmaroSeoaneSopuerta2014,AntogniniEtAl2014}.

The initial conditions for all experiments are a binary of two HSBs plus an
unbound interloper.  We note that during the interactions, there are long-lived
hierarchical triples forming, which are hence also included in the experiments.
We assume a relative velocity corresponding to the velocity dispersion of a
NSC, $50\,{\rm km\,s}^{-1}$.  Initially the code calculates $b_0$, the impact
parameter corresponding to a classical, point-mass closest approach of $2\,a$.
It then performs scattering experiments with $b$ increasing from near zero,
following a distribution uniform in area, $dN/db \propto b$.  It continues
performing classical, point-mass closest approach of $2\,a_0$, where $a_0$ is
the initial semimajor axis of the HSB binary.  It then performs scattering
experiments with $b$ increasing from near zero.  It continues performing
experiments as long as $b$ is less than $b_0$, or b is less than two times the
last $b$ for which the encounter was either a recordable event or a resonant
encounter. The reason for the latter constraint on $b$ is to ensure that any
interesting encounters that might occur at large $b$ are not missed.   The
masses of the HSBs are set to $m_1=m_2=30\,M_{\odot}$.  In a dense stellar
environment such as a NSC, a BH binary becomes hard when its kinetic energy per
unit mass, $G\mu/(2a_0)$ (with $\mu=m_1m_2/(m_1+m_2)$ the reduced mass of the
binary), becomes greater than the kinetic energy per unit mass of the field
stars, $3 \sigma^2 /2$.  This definition of a hard binary gives a critical
semi-major axis, $a_H = G\mu/(3\sigma^2)$, which is {about $1.8$ AU}
when $m_1=m_2= 30~M_\odot$ and $\sigma=50~{\rm km~s^{-1}}$.  A genuinely hard
binary (i.e.  one with binding energy $E_{\rm bin}=10\,kT$, with $kT\sim3
\sigma^2 /2$ the average thermal energy of stars in the system, see e.g.
\citealt{HH2003}) should have a semi-major axis {$a_0$ about ten times
smaller} than the above critical value.  Hence, for the simulations we consider
only $a_0\le 0.1$ AU.

For each binary we assume a single encounter with an interloper. The reason
why we do not consider successive encounters can be seen from
Eq.~(\ref{eq.Gamma}): The encounter rate for a HSB binary (of total mass
$60\,M_{\odot}$ and a semimajor axis of 0.1 AU) with another star is $\sim
5\cdot10^{-9}~{\rm yr}^{-1}$, while the rate to interact with an HSB interloper
is $\sim 5\cdot 10^{-12}~{\rm yr}^{-1}$. I.e. for a Hubble time the binary will
interact with 2 stars and with no other HSB.

To first order, the initial mass function (IMF) can be approximated by two
well-separated mass scales. We hence consider interlopers of masses
$1\,M_\odot$ (representing within order or magnitude main sequence stars, white
dwarfs and neutron stars) and $10\,M_\odot$ (stellar-mass black holes).  The
relative abundance (i.e. the number fraction) of objects in these mass ranges
{after a relaxation time} is dominated by the lighter stars
\citep{Alexander2005}, and hence we use a {number} fraction of heavy mass
particles of $f_H= 10^{-3} \times f_L$, with $f_L$ the abundance of light
stars.

We ran $10^4$ samples for each of the experiments to do a statistical study.
The code scans different orbital parameters, such as the initial relative
velocity and impact parameter, and integrates the system until it has been
dynamically solved. This means that the three-body interaction has finished and
we are then left with either a new binary with new orbital elements and one of
the {particles} has been ejected, or the binary has coalesced during the
three-body interaction because it was set on a very radial orbit and the
pericenter was smaller than the semi-major axis defined for coalescence (see
below).

We then filter the results looking for binaries that are already in the
detector band. For instance, for aLIGO this means that we look for systems with
an orbital period of $P_{\rm orb} \leq 20$ seconds---so that the gravitational
wave frequency (the double of the orbital frequency), should be $f_{\rm GW}
\geq 10~{\rm Hz}$. For those systems we calculate the eccentricities at both
detector-entrance and coalescence. We assume as a first-order approximation
that the systems coalesce if $a= a_{\rm mrg} := 3 R_{\rm Schw}$, where $R_{\rm
Schw}$ is the Schwarzschild radius (for instance, for a $30\,M_{\odot}$,
$a_{\rm mrg} \sim 1.8\cdot10^{-6}~{\rm AU}$).

Even if from a dynamical point of view the three-body interaction has finished
before the binary enters the detector band, we want to know how many will be in
the detector band within a Hubble time.  For this, we evolve the orbital
parameters from the last snapshot in the evolution that we have from the
numerical code, and we evolve them with an approximation of Keplerian orbits,
as in \cite{Peters64}.

The equations evolve a system under the assumption that gravitational radiation
is the only source of shrinkage of the semi-major axis of the binary, which is
valid in our situation, since we do not allow the binary to interact with any
other interloper after the first three-body scattering process.  We then evolve
them until they reach $f_{\rm GW} \leq 10~{\rm Hz}$, provided this happens in
within a Hubble time. Of those that enter into the detector band, we record the
eccentricity.

The calculations were halted whenever the orbital speed of binary reached $\sim
30\%$ c (the speed of light), since the $\PN$ expansion is not valid for higher
velocities. Furthermore, once that point in the evolution is reached, the
binary will always coalesce. The description of the parameter space of the
simulations given below corresponds to that point, which we refer to hereafter
as the {\em merger point} in the evolution of the binary, even if it does not
strictly corresponds to the real coalescence of the objects.  To speed up
calculations, the code analytically treats weakly tidally perturbed binaries:
{Any hierarchies are analytically treated whenever they are tidally
perturbed less than the tidal perturbation tolerance. This means that in a
hierarchical triple in which the eccentricity of the outer binary is very
large, the inner binary could be treated analytically at each apocenter passage
in the orbit---provided it is not strongly tidally perturbed \citep[for more details, see section 3.3.4 of][]{FregeauEtAl04}.
} To make this feature work with $\PN$ gravity, we added the proviso that a
weakly tidally perturbed binary must also have orbital speed at pericenter
smaller than $\alpha\cdot c$ to be treated analytically.  We typically take
$\alpha=0.05$, which provides a good compromise between accuracy and
computational speed.

\begin{figure*}
\resizebox{\hsize}{!}
          {\includegraphics[scale=1,clip]{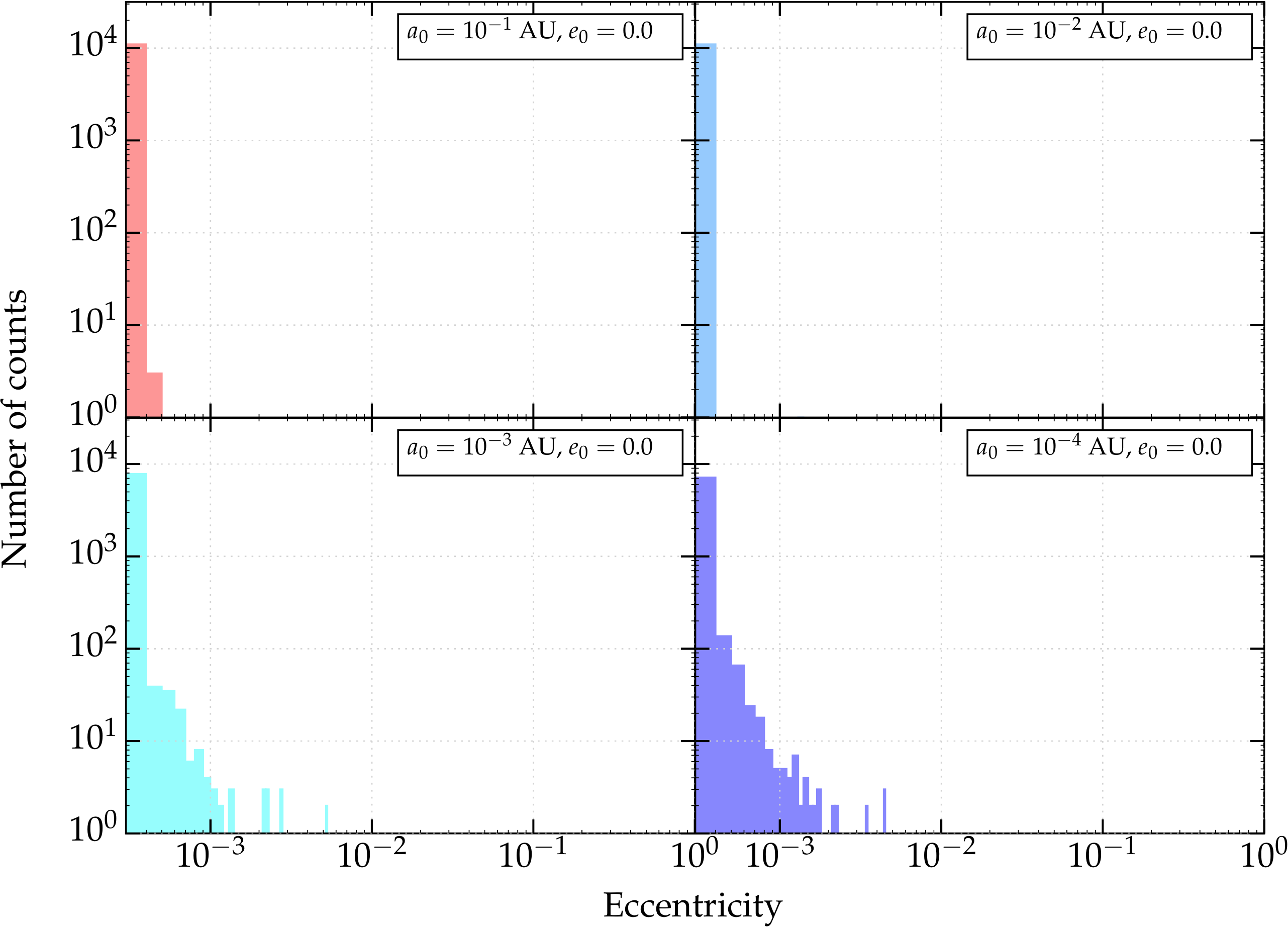}}
\caption
   {
Distribution of eccentricities for the binaries described in the text. From the
left to the right, top to bottom we show the distribution at detector entrance
(aLIGO) for a $10^{-1},\,10^{-2},\,10^{-3}~{\rm and}\,10^{-4}\,{\rm AU}$ binary
which initially was perfectly circular after interacting with stars in the NSC
(for the choice of parameters, see text).
%The last two panels, respectively,
%the distribution for a $10^{-4}\,{\rm AU}$ binary that initially had an
%eccentricity of $e_0 = 0.9$ at detector entrance (upper panel) and just before
%coalescence (lower panel, see text).
The NSC has a velocity dispersion of
$50\,{\rm km\,s}^{-1}$.
   }
\label{fig.hist_circular}
\end{figure*}

\begin{figure*}
\resizebox{\hsize}{!}
          {\includegraphics[scale=1,clip]{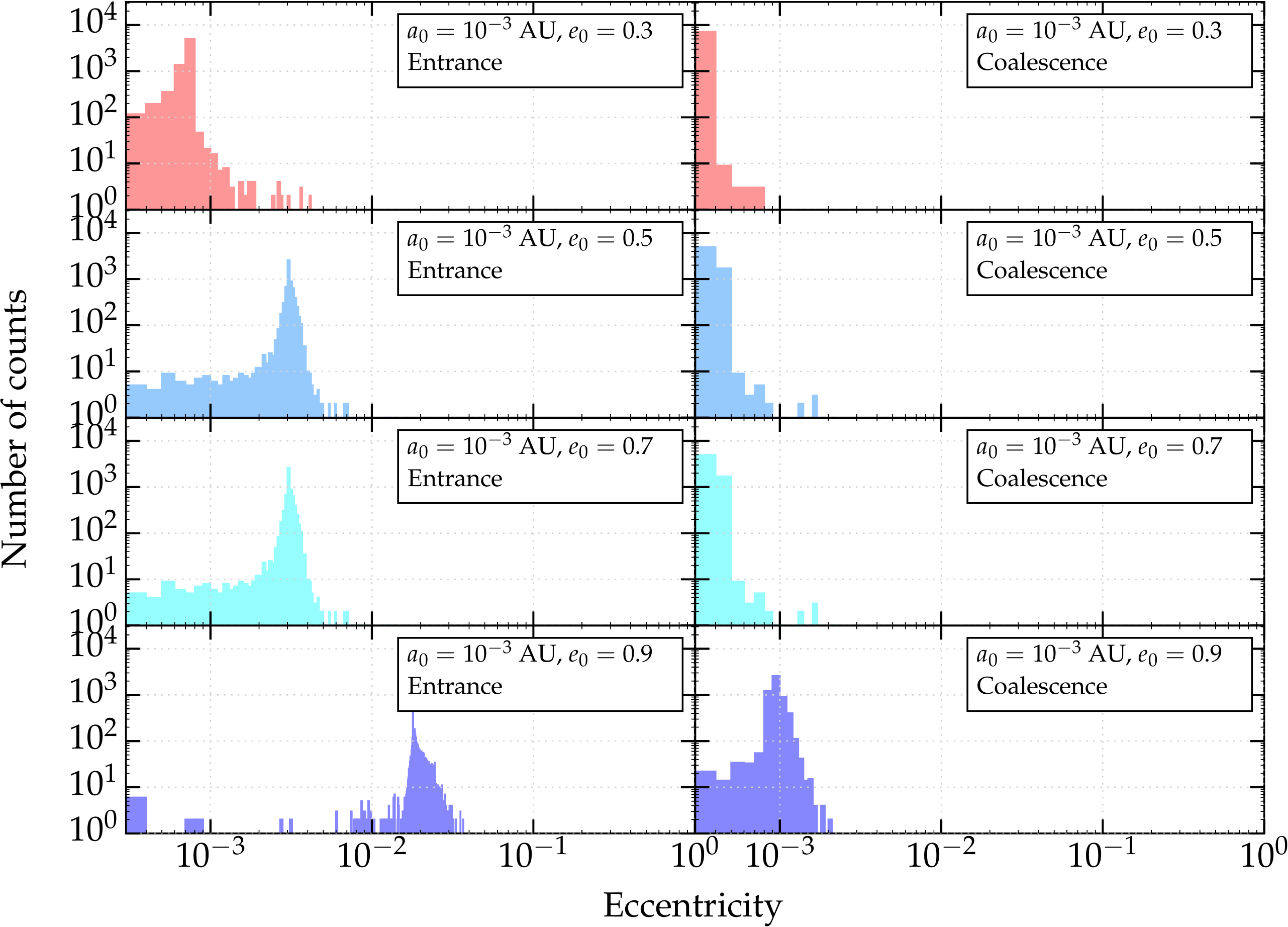}}
\caption
   {
Same as Fig.~(\ref{fig.hist_circular}) but for a $10^{-3}\,{\rm AU}$ binary and
four different inital eccentricities of 0.3, 0.5, 0.7 and 0.9 from the top to the
bottom, at detector entrance (left panels) and just before coalescence (right panels),
see text.
   }
\label{fig.hist_ecc_a1e-3}
\end{figure*}

\begin{figure*}
\resizebox{\hsize}{!}
          {\includegraphics[scale=1,clip]{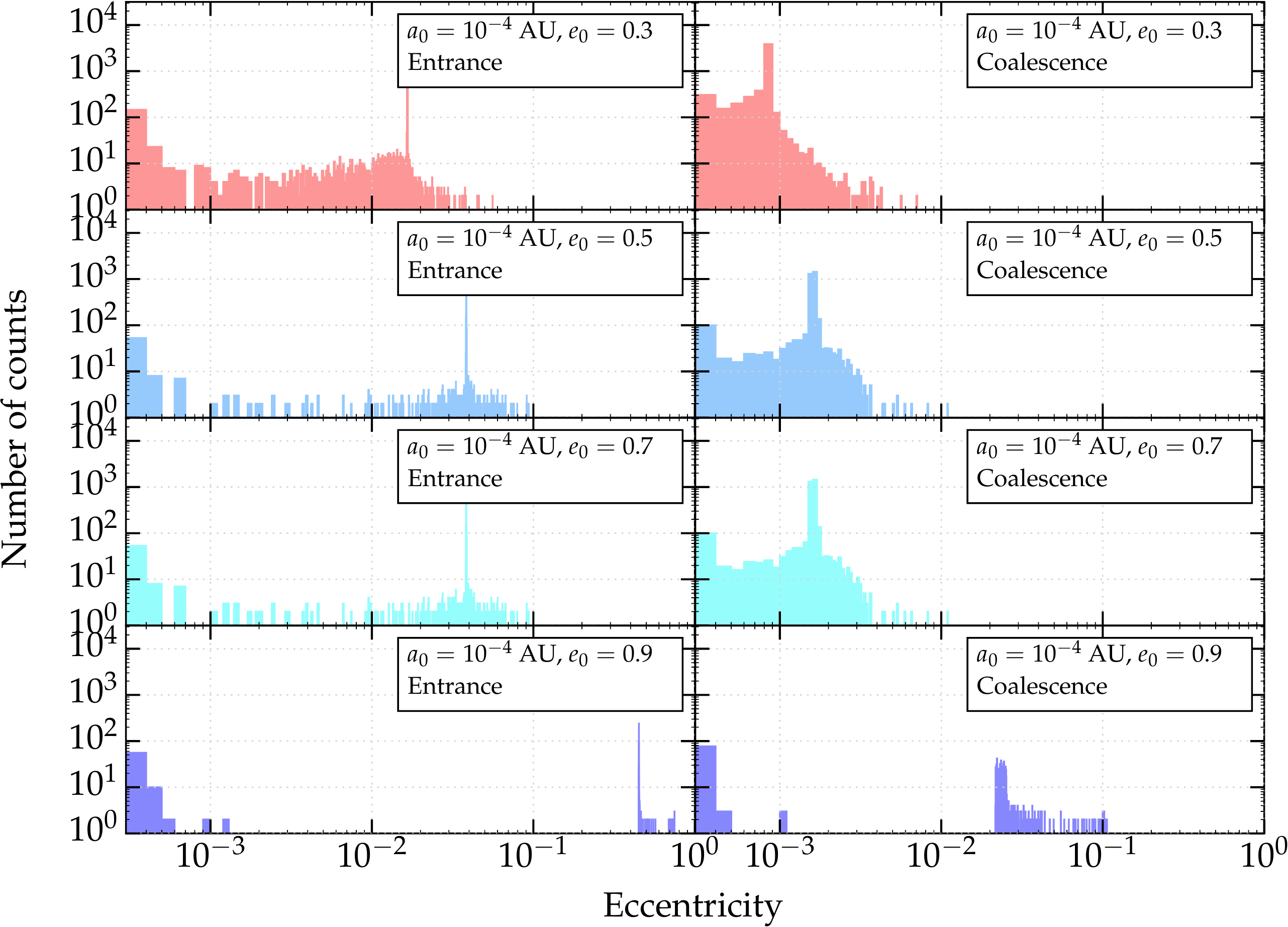}}
\caption
   {
Same as Fig.~(\ref{fig.hist_ecc_a1e-3}) but for a $10^{-4}\,{\rm AU}$ binary.
   }
\label{fig.hist_ecc_a1e-4}
\end{figure*}

In Fig.~\ref{fig.hist_circular} we display the distribution of eccentricities
for the scattering experiments {at detector entrance}.  We can readily
see that there is a trend towards higher eccentricites as the binary is tighter
because the tighter the binary, the shorter the time to coalesce, and the
shorter time for GW radiation to reduce the eccentricity. We also explore
binaries that initially are more eccentric, as displayed in
Figs.~(\ref{fig.hist_ecc_a1e-3}, \ref{fig.hist_ecc_a1e-4}), initially starting
with eccentricities of 0.3, 0.5, 0.7, and 0.9. For these, only a negligible
fraction of the experiments with a high initial eccentricity and a very tight
semimajor axis achieves a final eccentricity of about $\sim 0.1$ and a very few
events (about $3\%$ of the binaries with initial $e_0=0.9$ which are very
tight) enter the detector band with $e \sim 0.5--0.7$, to completely
circularise a bit before coalescence. All other results stay well below $\sim
0.1$. The relative fraction of eccentric binaries is so low that they can be
ignored even for the overoptimistic upper limit cases.

Current search strategies are based on templates of circular binaries, so that
the size of the parameter space will be manageable. Although this blinds them
to most eccentric mergers \citep{HuertaBrown2013,Favata2014}, and in principle
to the richer information contained in the gravitational wave signals
associated with those mergers, these events are so rare that they can be safely
ignored. The same applies to field binaries because, as explained before, they
will also be circular.

\section{Discussion and Conclusions}

In this work we have addressed the formation channels for BH binaries in the
range of observation of ground-based detectors. We find that in the two cases
(binaries forming isolated or via dynamical interactions) the most likely is
that the binary members are two BHs in the the mass range of $30-100~M_\odot$,
which we call ``hyperstellar black holes'', produced by massive WR stars.

We show that the two binary members, again in the two different formation
channels, will be slow rotators, meaning a spin magnitude of $a<0.5$. We show
that there is (so far) no possible physical process capable of spinning a HSB
up to values $a>0.5$ in a binary.

We carry out a statical study of the eccentricity distribution for a binary of
two HSBs in a dense stellar environment. For this we use a numerical Newtonian
code in which we have implemented relativistic correcting terms. We prove that
HSB binaries will predominantely have almost circular orbits at detector
entrance, so that eccentric binaries can be ignored in the searches.  Field BH
binaries formed in isolation have been shown to also have almost circular
orbits at aLIGO frequencies.

Moreover, these binaries should have similar masses. In
Section~\ref{sec:cluster} we gave reasons based in relaxation and we referenced
as well the work of \cite{MillerLauburg09} for the case of dynamical binaries. Field
binaries will also very likely have similar masses, as has been shown with
binary evolution and population synthesis codes \citep{demink15,dominik15}.

We therefore predict that HSB binaries will be observed by ground-based
detectors with similar masses, low spin magnitudes and almost zero
eccentricities, regardless of where they have formed. This has the potential to speed up the searches.

Moreover, a binary of two $30\,M_{\odot}$ BHs, i.e. a HSB binary, will be seen
at farther distance than a binary of two $10\,M_{\odot}$ BHs. So the
increased volume will ensure that many more massive HSBs will be detected.

Although there is so far no evidence for the existence of HSBs, we deem the
non-detection to be related to the limitation of conventional observations in
the electromagnetic wavebands. Stellar-origin BHs are discovered whenever a BH
is accreting material from a companion star \citep{mcclintock14}. For HSBs, the
lifespan in such a configuration is short: The companion star is very likely
also a WR star, since they should have similar masses, as we have seen. Since
the lifetime of a WR star is only 5 Myr, the time window to observe a HSB
binary is very short (even shorter if the two WR stars collapse each to a HSB
simultaneously). Hence, HSBs stay virtually always dark in the electromagnetic
domain.

With the first detections being imminent, soon the symbiotic relation between
astrophysics and data analysis that we mentioned in the introduction will be
fulfilled. The upcoming detections will either confirm our prediction or rule
it out, and we will hence obtain information about the birth and evolution of
stellar black holes, as well as { about} their environments---information that is
virtually unaccessible to our old friend the photon.

\acknowledgments

We thank Craig Heinke for discussions on observational aspects related to the
determination of the spin, and Douglas Heggie for discussions about binaries.
We are thankful to David J. Vanecek for his advice and assistance in C, and
Cristi{\'a}n Maureira-Fredes for his help with python.  We are indebted with
Bruce Allen for granting access to the Atlas cluster, where the simulations
were performed, and to Carsten Aulbert for his administration. XC is supported
by CONICYT-Chile through Anillo (ACT1101). We thank Matthew Benacquista for
comments on the manuscript. PAS is thankful to Ladislav \v{S}ubr and the
Astronomical Institute of the Charles University in Prague for a visit, in
which this paper was finished.

\label{lastpage}
\end{document}